# Bowel Incision Closure with a Semi-Automated Robot-Assisted Laser Tissue Soldering System

Shani Arusi[1]†, Max Platkov[2]†, Barak Rosenberg[1], Svetlana Basov[3], Ido Ashbell[4], Tom Polovin[4], Yoel Chocron[5], Abraham Katzir[6], Ilana Nisky[1] and Uri Netz[4,5,*]

[1]Department of Biomedical Engineering, Ben-Gurion University of the Negev, Beer-Sheva, Israel.

[2]Nuclear Research Center Negev, Beer-Sheva, Israel.

[3]Department of Biomedical Engineering, Tel Aviv University, Tel Aviv, Israel.

[4]Faculty of Health Sciences, Ben Gurion University, Beer-Sheva, Israel.

[5]Department of Surgery A, Soroka University Medical Center, Beer-Sheva, Israel.

[6]School of Physics & Astronomy, Tel Aviv University, Tel Aviv, Israel.

†Equal Contribution.

*Corresponding Author: Uri Netz (urinetz@gmail.com)

**One Sentence Summary**

A Heineke-Mikulicz bowel incision closure was performed with a novel Semi-Autonomous Robot-Assisted Laser Tissue Soldering system.

**Keywords**

Robotic soldering, Laser soldering, Bowel closure, In-vivo.

**Abstract**

Traditional methods for closing gastrointestinal (GI) surgery incisions, like suturing and stapling, present significant challenges, including potentially life-threatening leaks. These techniques, especially in robot-assisted minimally invasive surgery (RAMIS), require advanced manual skills. While their repetitive and time-consuming nature makes them suitable candidates for automation, the automation process is complicated by the need for extensive contact with the tissue. Addressing this, we demonstrate a semi-autonomous contactless surgical procedure using our novel Robot-assisted Laser Tissue Soldering (RLTS) system on a live porcine bowel. Towards this in-vivo demonstration, we optimized soldering protocols and system parameters in ex-vivo experiments on porcine bowels and a porcine cadaver. To assess the RLTS system performance, we compared the pressure at which the anastomosis leaked between our robotic soldering and manual suturing. With the best setup, we advanced to an in-vivo Heineke Mikulicz closure on small bowel incision in live pigs and evaluated their healing for two weeks. All pigs successfully completing the procedure (N=5) survived without leaks and the histology indicated mucosal regeneration and fibrous tissue adhesion. This marks the first in-vivo semi-automated contactless incision closure, paving the way for automating GI surgery incision closure which has the potential to become an alternative to traditional methods.

# INTRODUCTION

Resecting or incising tissues in the gastrointestinal (GI) tract is frequently performed in surgeries such as tumor removal, bariatric, and trauma procedures. Currently, the common methods for closing incisions are sutures or staples. However, both techniques exert tension and compression forces and can potentially cause tissue damage. Improper placement or extensive tension can impair the healing process, and their presence can trigger an inflammatory reaction, which can lead to leaks. Other common risk factors for leaks include impaired anastomotic blood supply, smoking, previous chemo-radiation, intraoperative hypotension, and, recently discovered, specific bacterial strains in the GI flora shown to inhibit stroma regeneration[1]. Leaks in these situations are a major cause of concern which can result in additional operations, lengthened hospitalization period, and increased mortality rates. In addition, leaks lead to a significant financial burden[2].

Minimally invasive methods offer a shorter and easier recovery and better aesthetic outcomes as an alternative to traditional open surgery. These methods include traditional laparoscopic surgery, and robot-assisted minimally-invasive surgery (RAMIS)[3]. In RAMIS, the surgeon moves robotic manipulators to control surgical instruments inside the body of the patient. RAMIS may improve surgical outcomes over open and traditional laparoscopic surgery by decreasing the patient's recovery time, which is associated with less hospitalization time and diminished blood loss [4, 5] It is also advantageous for surgeons, as it provides smoother control of the tools by reducing natural hand tremor with more degrees of freedom, a higher range of movement, and superior ergonomics [6]. However, many clinical fields still do not use RAMIS systems due to the lack of haptic feedback. The surgeon can see the patient's tissue in a 3D high-definition visual system but cannot feel the tissue. Therefore, tasks such as suturing can become difficult as the surgeon may apply inappropriate force on the suture or tissues. Although the learning process during RAMIS is faster than traditional laparoscopic surgery [7], advanced training is required with both methods to suture efficiently and minimize leakage [8], [9].

As an alternative, many researchers are exploring the use of robotic systems for autonomous surgery[10-13], which eliminates the need for a surgeon's expertise. This approach aims for consistent procedures and patient safety, which are the ultimate objectives of any operation [14]. Due to its repetitive nature and being a time-consuming task, suturing is ideal for automation [11, 15, 16]. However, it requires direct contact with the tissue which can be a problem in a dynamic surgical environment with perturbations from breathing, heart contractions, and peristaltic bowel movements. Autonomous suturing has been researched and attempted, but it still requires advanced imaging solutions [11] and some level of supervision [17]. Moreover, autonomous suturing may still suffer from complications that are related to the sutures themselves.

Laser Tissue Soldering (LTS) offers a promising alternative to suturing and stapling for wound closure. In LTS a laser heats a biocompatible solder made of proteins such as albumin until it firmly crosslinks with tissue. This process is enhanced by adding a chromophore dye, such as Indocyanine green (ICG), which absorbs laser light to focus heat, with the penetration depth influenced by laser wavelength and dye concentration. According to the Beer-Lambert law, higher ICG concentrations lower the penetration depth. Aiming at optimizing the solder, studies examined various materials and concentrations, indicating a tissue-specific interaction, with bovine serum albumin and ICG as common choices [18-21]. Findings suggest that high albumin concentrations strengthen tissue fusion, and lower ICG concentrations deepen laser penetration, enhancing tensile strength. In-vivo LTS studies showed high efficiency on soft, thin tissues such as skin [22], blood vessels [23], and nerves [24]. However, tissue thickness also plays a role in the strength of bonding, and the thicker tissues of the gastrointestinal tract

[25], [26] require higher laser penetration depth. In our review paper [21] we found that only one out of 19 studies that performed LTS for tissue bonding in the small intestine was an in-vivo study, and it was on rabbits [27]. Their histological results after two weeks of healing showed that the LTS did not cause thermal injury, and resulted in improved mucosal re-epithelization, reduced inflammation, and less bowel narrowing compared to suturing. However, LTS's applicability in human GI surgeries necessitates in-vivo testing on thick tissues such as porcine bowels, which were not demonstrated to date. Tight control of heating dose and trajectory is critical to ensure that the solder bonds effectively without causing collateral damage to surrounding tissues. As a result, the success of such manual LTS relies heavily on the manual skill of an operator. Hence, in LTS, just like in suturing or stapling, an automated alternative could provide substantial benefit to patients.

To bridge this gap and revolutionize surgical procedures, we aimed to provide a reliable, automated, and contactless alternative to traditional suturing based on the advantages of RAMIS and LTS: the Robot Laser Tissue Soldering (RLTS). In our previous work, we demonstrated the feasibility of this approach on a mouse skin model [28], where the temperature control system allowed a constant temperature profile, and the robotic system allowed a precise trajectory, together achieving strong bonding. Despite achieving promising outcomes with a mouse skin model, the capability of our system to handle the complexities of thicker, multilayered tissues, like those in the intestines of large animals, is yet to be developed. Additionally, the procedure and the recovery process in large live animals must be carefully evaluated to ensure that the bonded area heals properly, maintaining integrity under the mechanical stress of gastrointestinal movements.

In response to these challenges, in this paper we demonstrate a semi-autonomous contactless surgical procedure using our novel Robot-assisted Laser Tissue Soldering (RLTS) system on a live porcine bowel. This required fine-tuning various parameters, including robotic trajectories and the concentrations of solder and dye, to refine and enhance the system's performance. Our investigation commenced with a series of ex-vivo experiments on porcine intestines to determine the optimal protocol that achieves closure efficacy on par with or surpassing traditional suturing. Following this optimization, the refined protocol was applied in a surgical setting, initially on a cadaver and then on live, anesthetized animals. These animals were monitored for two weeks post-procedure to assess the healing outcomes, providing critical insights into the practical application and effectiveness of our RLTS system in a clinically relevant context.

# RESULTS

First, we performed a series of ex-vivo experiments to assess various aspects of the Robotic Laser Tissue Soldering (RLTS) system. The RLTS system integrates a laser for contactless tissue bonding under temperature control and a robotic component for precise movement control (Figure 1). These experiments focused on evaluating the impact of various parameters of this system, such as movement speed, proximity to tissue, and the concentrations of solder and dye on the quality of soldering. We assessed the results for each experiment by measuring the burst pressure: manual sutured and robotic soldered, against the median burst pressure of a control group of intact tissue from the same batch of bowels on the same day. Normalizing burst pressure values allowed for comparison across different experimental sessions and tissue samples. For the robotic soldered group, we performed a longitudinal incision in each piece of porcine intestine and applied two approximation sutures at the ends of the incision. The solder solution was manually applied to the incision area, and then soldering was performed. For the manual sutured group, we executed a standard single layer continuous suturing technique on identical incisions. After soldering or suturing, both sides of the intestinal sample were clamped, water was introduced into the cavity at a constant rate to steadily increase pressure, and the burst pressure was recorded. This measure serves as an ex-vivo proxy for bond quality, acknowledging that it doesn't fully predict the long-term integrity of the closure or potential complications like tissue necrosis, which could emerge from overly tight sutures or staples. Our goal was not necessarily to surpass the manual suturing method in burst pressure outcomes but to ensure that robotic soldering remains competitive in terms of bond strength by choosing the best parameters at each step. In addition, we prioritized identifying best protocol parameters to advance to in-vivo testing rather than achieving statistical differences between protocols. After choosing the best protocol parameters on simple incisions, we performed an experiment on a clinically relevant medical procedure, the Heineke-Mikulicz (HM) procedure. After achieving successful reliable closure in the ex-vivo intestine in the lab, we moved our setup to an animal testing facility and repeated the HM procedure on a fresh cadaver. Finally, we performed in-vivo HM procedures on the intestines of six live pigs. We monitored their healing process over a two-week period and completed the experiment with their sacrificing for histology analysis.

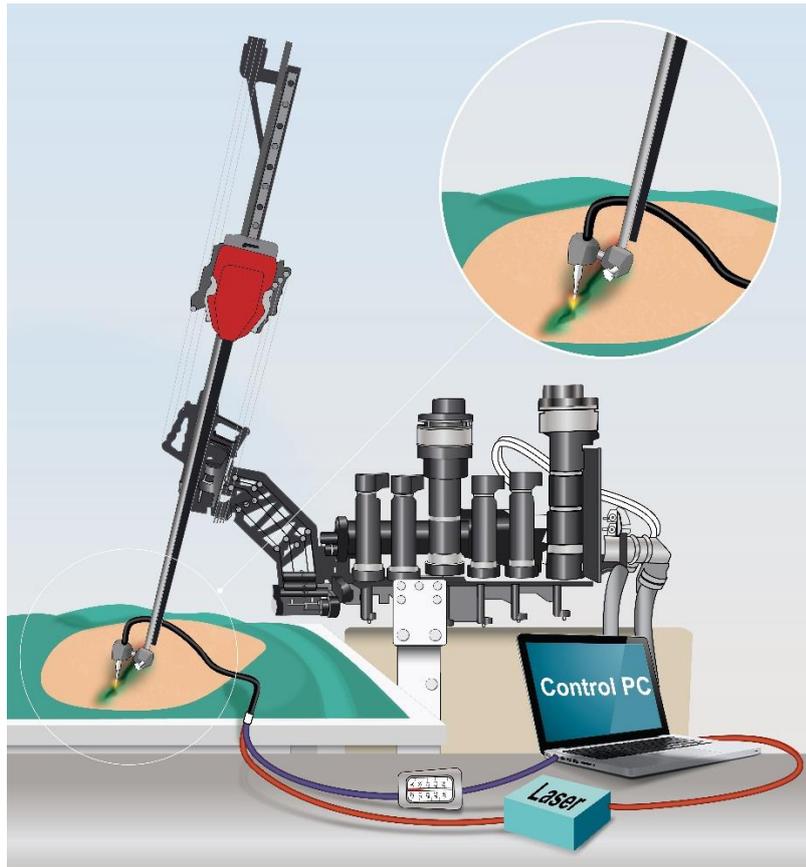

*Figure 1 : Illustration of the Robot Laser Tissue Soldering system.* *The system set up for the ex-vivo experiments. There are five fibers connected through a common mechanical fixture to the robotic tool: a single fiber providing laser illumination and four fibers for collection of the middle infrared radiation emanating from the hot sample, to be analyzed by the radiometer. The radiometer translates the middle-infrared signal to voltage and transmits it to a computer (connected with the blue wires) which operates a software feedback loop that drives the laser intensity to maintain a pre-calibrated temperature setting. The robotic tool is attached to the robotic arm of the Raven-II robotic system. This robotic system has a separate control system controlling the movement of the arm. The tissue was positioned under the laser and the robotic tooltip. The working distance can be changed with the moving stage beneath or by the robot movement along the Z axis (up or down).*

### *Ex-vivo experiments on linear incisions*

Our first step was to find the best robotic movement trajectory among the zig-zag, fast discrete, slow discrete and double slow discrete trajectories. In our first four experiments, we found that the burst pressure in robotic soldered specimens was lower than the manual-sutured results except for the double discrete movement trajectory (Figure 2A). For these four experimental protocols the albumin concentration was 800 mg\ml and the dye concentration was 0.6 mg\ml. In the first experiment, we examined the robotic movement with the zigzag trajectory. During this movement, the robot stopped for 12 seconds at each step in a structured track along the incision. Each laser beam movement was defined as two steps to the right, a step forward, two steps to the left, another step forward, and so on. In this case, the laser beam was narrow since the laser was close to the tissue at a distance of 7 mm. This movement allowed us to cover the entire incision area (Figure 2B). For this protocol, there was a large variation between the robotic soldered samples, and the overall result was inferior to the manual suturing group. In the second experiment, a fast discrete trajectory was performed in a straight line at a velocity of 0.3 cm/sec with a 10-second dwell time with a distance from the tissue of 35mm. The results for the solder group were again poor. However, for the same movement at a lower velocity (of 0.1 cm/sec), we achieved a higher burst pressure (Figures 2A and 2C) but still inferior to suturing. Therefore, in the fourth experiment, we duplicated the exposure time on the incision area. This approach entailed repeating the linear movement twice, with the robot executing a straight-line back-and-forth motion at a velocity of 0.1 cm/sec, pausing for 10 seconds at each interval. When the double discrete movement trajectory was implemented, we found an advantage for robotic soldering over manual suturing (Figure 2A). Using a straight-line movement protocol at a larger distance from the tissue achieved a faster soldering with a wider laser beam less susceptible to imperfections in the tissue edges position (Figure 2D and Figure 3B). Statistically significant differences between protocols were not critical to our transition to the next step, but nevertheless, statistical analysis also supported the advantage of the best protocol over the worst with a two-way ANOVA model. We found a statistically significant effect of the incision closure technique (manual suturing or robotic soldering) factor ($F_{1,21} = 10.19, p = 0.0044$) and for the interaction between the incision closure technique and the robotic movement trajectory factors ($F_{3,21} = 4.09, p = 0.0196$), but not for the robotic movement trajectory (zigzag, fast, slow or double discrete trajectory) factor ($F_{3,21} = 2.78, p = 0.066$). A Post-hoc analysis showed that the double discrete movement trajectory was statistically significant better than the zig-zag movement trajectory ($t_{21} = 2.96, p = 0.0074$) but not from the fast ($t_{21} = 1.16, p = 0.259$) and the slow ($t_{21} = 055, p = 0.59$) discrete movement trajectories.

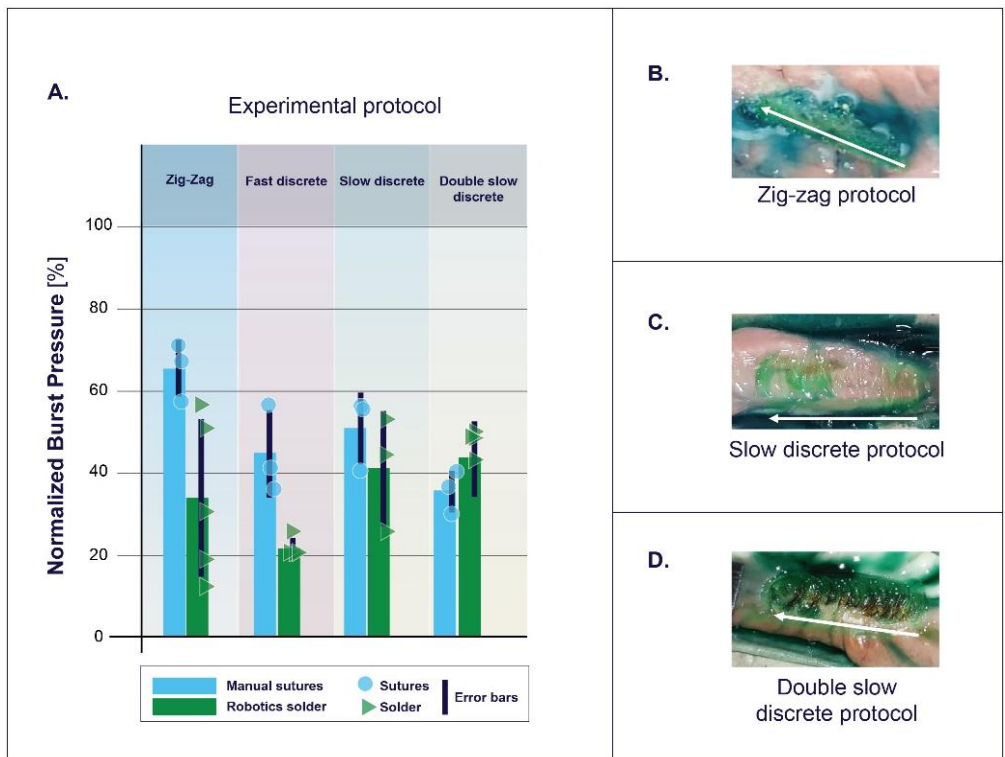

*Figure 2: Results for different robotic trajectories. A) The mean normalized burst pressure of manual suturing (light blue bar) and robotic soldering (dark green bar) for different robotic movement trajectories: the Zig Zag, the fast discrete, the slow discrete and the double discrete movement trajectories. The error bars represent the standard deviation across samples, and the circles and triangles are the results for each sample. (B-D) Images after robotic soldering for different robotic trajectories: B) Zigzag, C) slow discrete and D) double slow discrete robotic movements. The double discrete protocol is also depicted in Figure 3B. The white arrow is under the soldering area.*

We then imparted on refining the soldering parameters by examining the solder and dye concentrations. We began by maintaining a constant dye concentration of 0.6 mg/ml while experimenting with various solder concentrations. The best results were achieved with an albumin concentration of 800 mg/ml, which produced a burst pressure exceeding that of manually sutured samples, as shown in Figure 3A, even though the advantage was not statistically significant ($F_{3,13} = 0.93, p = 0.455$). The higher viscosity of the solder in this high albumin concentration facilitated its retention in the targeted area without spillage prior to soldering. This ability of the solder to stay in place justified the continuation of experiments using this albumin concentration. A visual assessment of tissue post-soldering using the double discrete trajectory, illustrated in Figure 3B, revealed a noticeable overlap in the laser-created circles, indicating a more comprehensive coverage than that achieved by the one-pass discrete trajectories, whether slow or fast, as documented in Figure 2C.

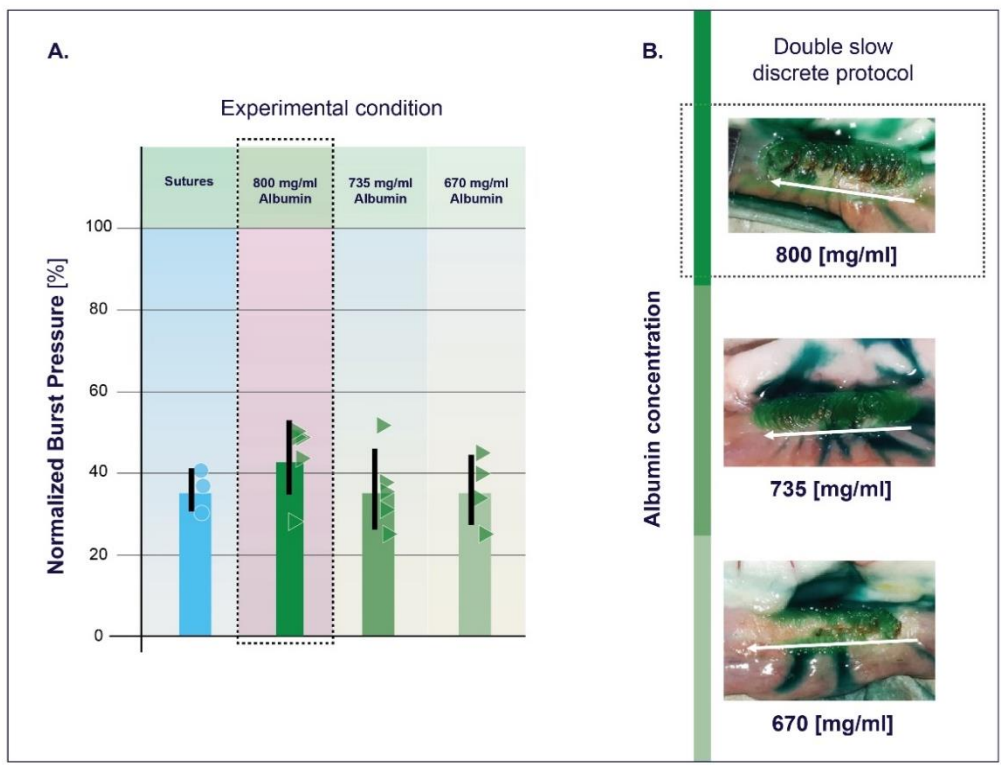

*Figure 3: Results for different solder concentrations. A) The mean normalized burst pressure for solder concentrations of 670, 735, and 800 mg/ml (shades of green) compared to the result of manual sutures performed on the same day (light blue). The results of the manual sutured group and the robotic soldering with 800 mg\ml albumin group also shown in figure 2A. B) Images of the serosa side of robotic soldered tissues for different albumin concentrations: the highest concentration of 800 mg/ml (top panel, the same as figure 2D), 735 mg/ml (middle panel), and 670 mg/ml (bottom panel). The white arrow is under the soldering area. The dashed gray frame surrounds the selected protocol for the next experimental protocol.*

When examining different dye concentrations while maintaining the solder concentration at a constant value of 800 mg/ml, we found that low ICG concentrations yielded better results (Figure 4A). Soldering with 0.3 or 0.45 mg/ml of ICG resulted in higher burst pressures than the 0.6 mg/ml ICG concentration. The results were comparable to those of the manual sutured group on that day. This was further supported by a one-way ANOVA test where we did not find a statistically significant effect for the experimental condition (manual suturing or the three different ICG concentrations for robotic soldering) factor ($F_{3,12} = 1.97, p =$

0.172). We also visually examined the serosa side of the specimen before the burst pressure test (Figure 4B) and the mucosa side after the burst pressure (Figure 4C). We found that the best fusion was in the 0.45 mg/ml ICG concentration soldered tissues, as demonstrated by strong bonding on both sides of the incision, detailed in Figure 4C's middle panel.

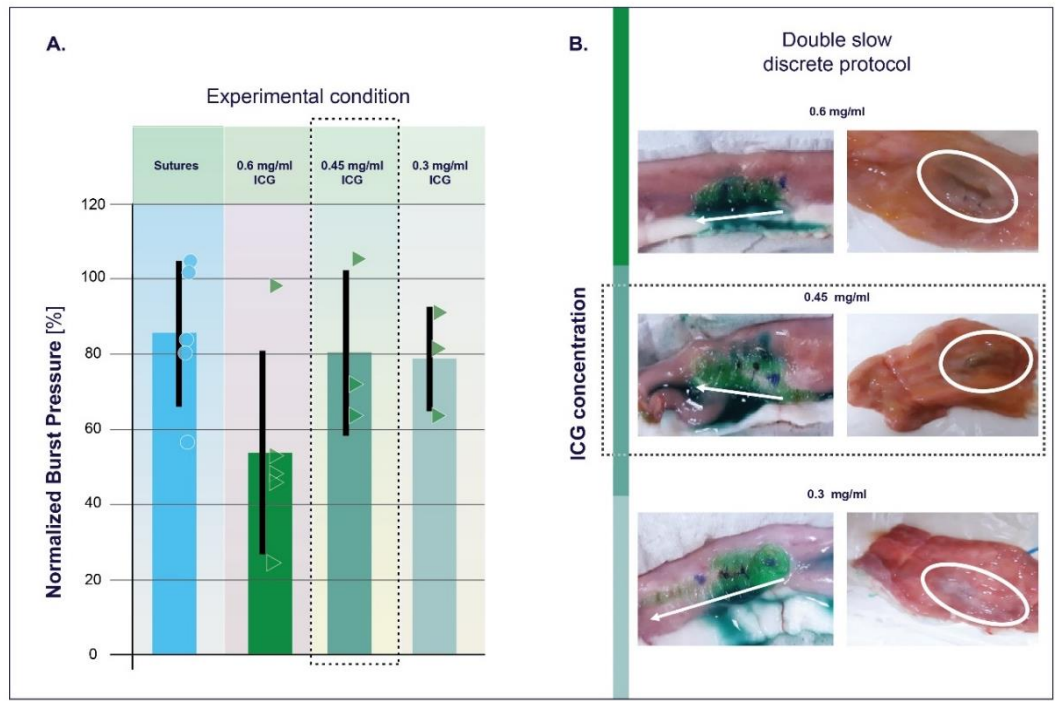

*Figure 4: Results for examination of dye concentrations.* A) The mean normalized burst pressure for ICG concentrations of 0.3, 0.45, and 0.6 mg/ml (shades of green) compared to the result of manual suturing group on the same day (light blue). The error bars represent the standard deviation for each group. B) Images of the tissues after robotic soldering from the serosa (left panel) and mucosa (right panel) side of the samples in experiment 5 for different ICG concentrations: the highest concentration of 0.6 mg/ml (top panel), 0.45 mg/ml (middle panel), and 0.3 mg/ml (bottom panel). On the left panel images the white arrow is under the soldering area and on the right panel images the white ellipses mark the soldering areas. The dashed gray frame surrounds the selected protocol for the next experimental protocol.

*Ex-vivo experiments with a Heineke-Mikulicz closure model*

After choosing the robotic movement parameters and the solder solution concentrations, we performed an experiment on a clinically relevant medical procedure, the Heineke-Mikulicz (HM) procedure. This procedure is used clinically to deal with strictures, and it entails a longitudinal incision along the intestine that is closed transversely, resulting in widening of the intestine at that section. To implement this, instead of only moving along the incision, we added a vertical movement, which created a plus-shaped pattern (Figure 5A). This model was successfully robotically soldered in our lab resulting in higher burst pressures than manual suturing (Figure 5B, leftmost pair of columns).

Following this success, we moved our setup to the institute for animal research of Kibbutz Lahav CRO and repeated the HM procedure on ex-vivo specimens and reproduced the results from the experiments in the lab at BGU. In both experimental facilities, we found that robotic soldering yielded superior results to sutures (Figure 5B left to the dashed line). Statistical analysis with a two-way ANOVA model also supported this conclusion. We found a statistically significant advantage for the robotic soldering over manual suturing ($F_{1,15} = 19.63, p = 0.0005$) but not for the experiment facility location (BGU lab vs Lahav CRO facility) factor ($F_{1,15} = 1.28, p = 0.276$) and their interaction ($F_{1,15} = 1.22, p = 0.287$). As a last step prior to in-vivo studies, we moved the system to the operating room and adjusted the settings of the system to meet the different environmental conditions in the operating room. We performed HM procedures on a fresh porcine cadaver and confirmed that the resulting burst pressure of the robotic soldered cadaver intestines are comparable to those achieved outside of the operating room (Figure 5B, right to the dashed line). The successful reproduction of these results on two different experimental days and on the cadaver confirmed that this protocol is durable for in-vivo experiments.

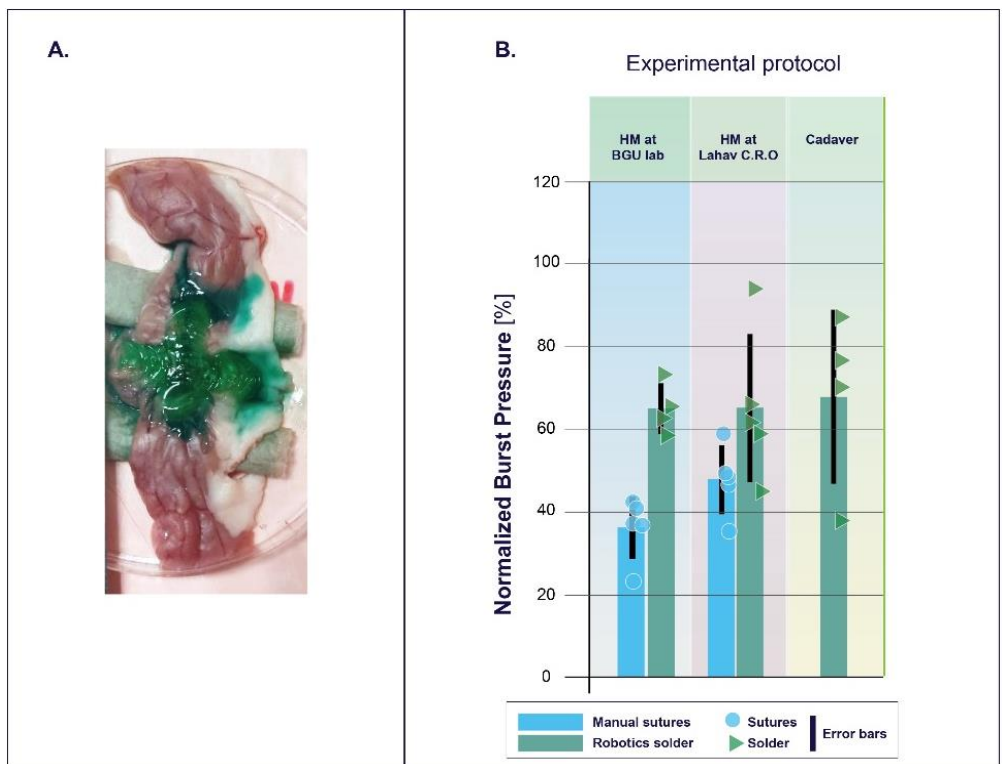

*Figure 5: Results for the Heineke-Mikulicz (HM) procedure in ex-vivo experiments. A) A tissue image after robotic soldering using the HM protocol. B) The mean normalized burst pressure for different experimental days when both manual suturing (light blue bar) and robotic soldering (dark green bar) were performed by the HM method. For the cadaver experiment, only robotic soldering was performed. The error bars represent the standard deviation, and the circles and triangles mark the results of each suturing and soldering sample relatively.*

*In-vivo experiment*

We performed robotic laser tissue soldering in-vivo on porcine small intestines using the HM procedure with our new protocol on a total of six pigs (weighting ~50-60 kg) (Figure 6A). One of the pigs died due to heart failure during anesthesia, an event unrelated to our experimental procedures. The other animals (N=5) were monitored for two weeks after surgery. All of them returned to normal intestinal function within one week. To evaluate the healing process after two weeks, we examined samples of the soldered area, which we marked by non- absorbable sutures (Figure 6B). We compared histology results in three different areas for each sample: control, mesenteric, and antimesenteric. For the antimesenteric area, where the soldering was performed, there was mucosal regeneration and serosal fibrosis (an example of the findings is shown in Figure 6C). The area of well-organized fibrous tissue (FT) was approximately 5.5 mm wide (for this specific tissue). This result indicates a successful recovery of the incision after robotic soldering. In the mesenteric area, at the opposite side of the soldering area, the remnants of our procedure, such as approximation sutures and soldering edges, were evident in some cases. Mucosal regeneration was seen in the previous incision areas. Importantly, across all examined sections, there were no indications of inflammatory response or thermal damage to the tissue, emphasizing the method's safety and effectiveness in facilitating tissue recovery.

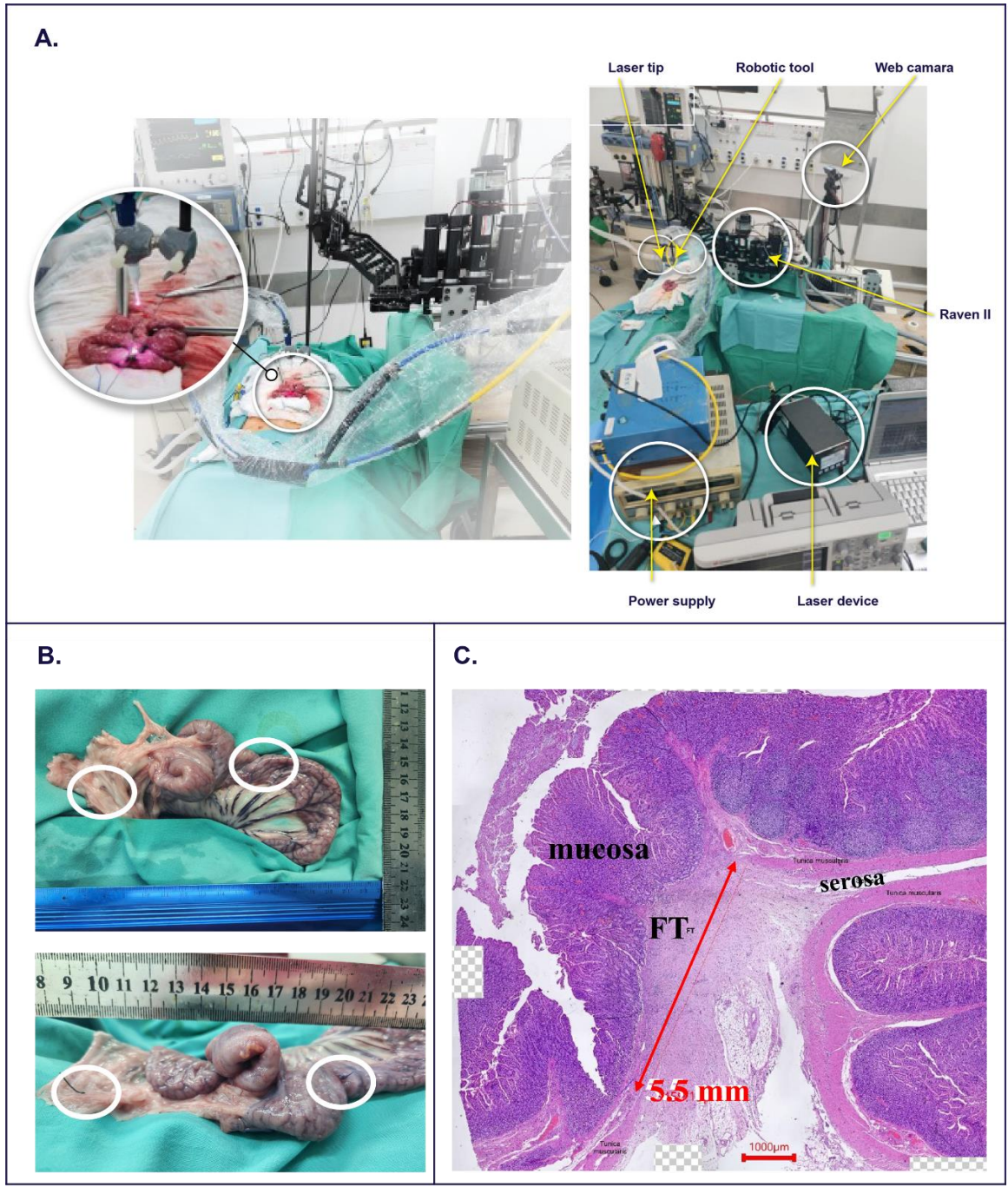

*Figure 6: **In vivo experiments:*** *A) The Robot Laser Tissue Soldering system in the Lahav CRO operation room for the in-vivo experiments. Here, only one web camera monitored the surgical area. (B-C) The results two weeks after the RLTS Heineke-Mikulicz procedure in the in-vivo experiments. B) Representative examples of the soldered tissue area after two weeks from two viewpoints of the same animal. The non-absorbable sutures are marked by the white circles, representing the beginning and the end of the soldering area. C) The histology results of the antimesenteric side of the tissue in A. The red arrow indicates the fibrous tissue (FT).*

# DISCUSSION

We developed and refined a novel robot-assisted laser tissue soldering system and demonstrated its effectiveness in closing incisions on the small intestines in a Heineke-Mikulicz procedure model. Through rigorous ex-vivo experimentation we focused on optimizing solder solution concentrations and robotic movement parameters to facilitate a semi-automatic, contactless procedure. Following optimization and streamlining of our system and experimental protocol, we proceeded to test it on the bowels of live pigs. In the ex-vivo experiments, we achieved comparable (and in some cases even higher) burst pressure to traditional manual suturing. In the in-vivo demonstration no leaks were seen and positive histological signs of healing were seen in all the five pigs at two weeks post-surgery. This is the first demonstration of robot-assisted laser tissue soldering on live pigs with healing assessment.

We demonstrated that a high concentration of bovine serum albumin (BSA) yielded the highest burst pressure, which was also higher than manual suturing. This result was also supported by McNally et al. [18, 29, 30], who found a higher tensile strength for the 60% BSA concentration and a poor result for the 25% BSA. Our choice of albumin concentrations was made according to previous literature on laser tissue soldering [21]. The highest value (800 mg/ml) was chosen and was also the highest one that we were able to produce. Although the advantage of this concentration in terms of burst pressure assessment was not statistically significant, we chose this concentration to attain a viscous solder that could be applied on the incision area without leaking onto the tissue sides.

Adding a chromophore to the solder solution results in concentrated heating at the location of the dye. Our experiments showed that low concentrations of ICG led to better tissue fusion throughout the bowel wall on both the inner (Mucosa) and outer (Serosa) sides as in McNally et al. [29, 31]. A possible explanation could be that high ICG concentrations completely absorbed the laser beam radiation, not permitting it to reach the deeper layers. By using a low concentration, we obtained a higher burst pressure of the soldered tissue than was found for the manual sutured group. Therefore, we chose the middle value tested, 0.45 mg/ml, for our protocol because we obtained a better fusion for the mucosa layer.

The RLTS system provides the benefit of a contactless procedure, eliminating the need for direct interaction with the tissue of the patient for closing the incision. We used a straight-line movement trajectory at a larger distance from the tissue which achieved a faster soldering with a wider laser beam less susceptible to imperfections and malposition of the tissue edges. Differing from the closure of incisions with supervised or autonomous suturing methods proposed in previous studies[12, 31], our semi-automated RLTS system uses a contactless method. Although their artificial intelligence (AI) control system for identifying the incision area deals with the problem of intestinal movement [11], extensive supervision is still required in dynamic environments. Our system compensates for such dynamics, including small peristaltic movements of the intestine, thanks to our integrated laser intensity adjustments guided by a precise closed-loop temperature control mechanism [28]. For compensating big peristaltic movements of the intestine further investigation need to be done to improve the robotic control system in the minimally invasive mode.

Despite these achievements, some limitations exist in the current implementation. This system still cannot replace the surgeons and technicians needed to monitor the system's performance. Our system autonomy level is 2, according to Yang et al. [17]. However, we aspire to make the entire system and protocol fully automated in the future by removing the human role from the control loop in several steps. First, we will improve the solder solution application onto the tissue inside the patient's body. Consequently, the surgeon will not need to place the material manually. This could be achieved by designing a unique tool fixed on a second robotic arm of the Raven robot, which will allow injection of the solder into the incision.

This solution will also enable minimally invasive soldering. Secondly, we plan to improve the monitoring of the incision area using image processing. We will program a vision-based trajectory tracking for the RLTS system that will allow for the implementation of complicated trajectories to achieve full anastomotic bonding.

In conclusion, our novel development and application of the Robot-assisted Laser Tissue Soldering (RLTS) system represents a significant advancement in the field of minimally invasive gastrointestinal surgery. Our findings demonstrate improved burst pressures and positive histological signs of healing compared to traditional manual suturing. Notably, the integration of a chromophore into the solder solution and the selection of optimal albumin concentrations were key to improving tissue fusion and maintaining the integrity of the soldered area. Despite the current system's level 2 autonomy, our future goals aim at fully automating the RLTS procedure, reducing the necessity for human intervention, and further enhancing the precision and reliability of incision closures. By advancing the RLTS technology and protocol, we are not only contributing to the broader application of laser tissue soldering systems but also paving the way for safer, more efficient surgical practices that could revolutionize patient care in gastrointestinal surgery.

## MATERIALS AND METHODS

*The experimental setup*
The Robot-Assisted Laser Tissue Soldering (RLTS) system is comprised of a patient-side manipulator (PM, Raven-II) that allows automated control procedures and a laser device that allows rapid soldering at a constant temperature (Figure 1).

*The laser system*
Our laser soldering system controls the temperature of the tissue with a proportional–integral–derivative (PID) controller, as described in [28]. In this closed-loop control, the laser heats the tissue and measures the temperature above it. Subsequently, the laser heating power dynamically changes to achieve constant temperature on the tissue. Our system consisted of five optical fibers packed in a common channel fixed to the robotic tool. There were two types of fibers: a transmitting visible radiation fiber for heating the tissue and a middle infrared radiation (mid-IR) fiber for temperature monitoring. This system has a laser with an 808 nm wavelength, which allows higher penetration depth. Therefore, temperature calibration was performed at the working area with thermal emitters with known temperatures between 20° and 60° [28, 31, 32].

The working area, defined as the distance between the fiber and the tissue during the experiment, varied between the robotic movement trajectories. Therefore, the Gaussian profile of the laser beam changed proportionally to this distance. As such, a larger spot has been created by increasing the distance between the laser and the tissue. The change in this distance affects the temperature applied to the tissue.

*The robotic system*
The Raven II is a research platform designed for surgical robotic applications[35]. We used the Raven II as our robotic system because it is an open platform in which the control system can be programmed to our needs. In this study, we used only the right arm of the Raven II. A custom-written software allows surgeons to control instruments that are mounted on the arms of the Raven by using a haptic device, keyboard control, or automatic open loop control. Here, we used the automatic open loop control. We used a MATLAB App, which allowed the user to enter the parameters for the desired movement path (according to the parameters explained for the different robotic movement trajectories). The movement was subsequently initiated by pressing on the relevant trajectory button in the App. The position of the end effector is updated in real time according to the planned trajectory. As a safety mechanism for stopping the robot movement in case of a problem, we also defined a termination button.

To adapt the Raven system to the laser fibers tip, a da Vinci Surgical tool (Intuitive Surgical, Sunnyvale, USA) was attached to the system using special adapters (Applied Dexterity, Seattle, USA). The adapter connected the Raven to the tool, as shown in [28], and then a custom-made fixture, connected the laser tip to the tool (Figure 6A).

*Robotic movement trajectories*
We examined different robotic movement trajectories: Zigzag, discrete, double discrete, and Heineke-Mikulicz (HM) trajectories (Figure 7). They differ by their path and velocity, while the exposure time at each point in the inaction area was maintained constant. We assumed a continuous, straight soldering path with a constant distance from the tissue. We could make this assumption due to the laser system PID controller compensation that corrected for the temperature.

(1) *Zigzag trajectory:* the movement was performed as a sequence of seven discrete steps, with the tip of the laser tool at a distance of 7 mm above the tissue. The robot made a sequence of two movements to the right, one forward, two left, then forward again, and so on (Figure 7A left panel). The distance between two points was 1 mm in the vertical direction and 1.5 mm in the horizontal direction, remaining for 12 sec of dwell time at each point.

(2) *Discrete trajectory:* the robot moved along the incision discretely, with a gap of 1.5 or 2 mm at a velocity of 0.1 mm/sec. At each point, the robot was held for 10 seconds of dwell time before moving on to the next step (Figure 7B).

(3) *Double discrete trajectory,* the robot moved back and forth in the same trajectory parameters as the discrete trajectory in (2) (Figure 7C).

(4) *Heineke-Mikulicz trajectory,* the robot moved along the incision area in a double discrete trajectory, with a gap of 2 mm at a velocity of 0.1 mm/sec, and then another movement was performed orthogonally with the same parameters (Figure 7D).

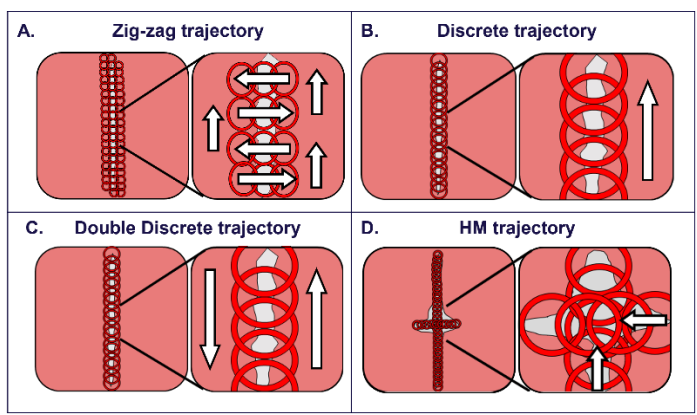

*Figure 7: Robotic movements trajectories. For covering the entire incision area using the laser beam, we tested several trajectories of the robot's movement: A) Zigzag, B) Discrete, C) Double discrete, and D) HM trajectories. The incision area (gray area) on the tissue (pink area) is covered by the laser beam (red circles). There are overlaps between the current beam hit and the previous hit area. The white arrows from bottom to top indicate the direction of movement.*

*The experimental protocol:*

*Ex-vivo experiments*

We performed a total of seven ex-vivo experiments, each examining different parameters. We used a fresh porcine small intestine provided by the Institute for Animal Research of Kibbutz Lahav CRO in Israel. The bowel was stored at $4^O$C for up to 24 hours until the experiment was performed. At the beginning of each experiment, the intestine was washed and then kept moist with normal saline solution at room temperature.

In each experiment, we compared three experiment groups: undamaged, manually sutured, and RLTS soldered tissues. The undamaged tissue, i.e., tissue without any incision, was used as a control group. For the sutured and soldered groups, we made a standardized longitudinal 10 mm incision in each 100 mm segment of the porcine intestine. For the manual suture group, one layer of continuous Vicryl 4.0 sutures was performed in an end-to-end fashion. For the soldered tissue, we applied isolated approximation sutures of Vicryl 4.0, two for the straight path movements and three for the HM procedure.

The solder solution, made from bovine serum albumin (BSA) and Indocyanine Green (ICG), was then manually applied to the incision area. Next, we placed the tissue under the tip of the RLTS system tools (Figure 1 and 6). The soldering path was a straight-line path according to the height from the tissue, the velocity, distance, and dwell time between points, as shown in Table 1. The temperature set point was set to $40^O$C. This setpoint was lower than the setpoint that we used in previous experiment with thinner tissues and weaker laser [28], and it is below the optimal heating temperature of 60 degrees that needs to be accumulated for at least 10 seconds for successful adhesion. However, pilot testing revealed that using a higher setpoint caused thermal damage to tissue, and hence we converged on this lower setpoint that allowed for successful adhesion. Note that this setpoint was calibrated with measurement of the radiation temperature from a black metal piece instead of the tissue, and we estimate that in the depth of the multilayered thick tissue of the intestine the needed temperature for adhesion accumulated. After soldering, we performed a burst pressure test.

The goal of the ex-vivo experiments was to optimize the protocol parameters. Therefore, in each experiment, we examined different parameters (Table 1): robotic movement trajectory, solder concentrations, dye concentrations, and incision closure (longitudinal vs transverse in HM). During these experiments, we performed various movements with the robot. First, we examined a protocol where the laser fiber was very close to the tissue (7 mm). In this case, the width of the laser beam was also small. A zigzag movement was required to cover the entire incision area. In the second protocol, the laser was set at 35 mm from the tissue, such that a faster straight-line path protocol was performed. We increased the dwell time between the different points to enable enough exposure time and examined different velocities. Finally, we examined a double exposure protocol repeating the entire movement backwards to increase laser exposure time. Using the chosen robotic movement trajectory, we examined different solder concentrations. The albumin solder was prepared from an aqueous solution of 670, 735, or 800 mg/ml bovine albumin mixed with 0.6 mg/ml of ICG. For the dye concentrations experiment, the albumin solder was prepared from an aqueous solution of 800 mg/ml bovine albumin mixed with 0.3, 0.45, or 0.6 mg/ml of ICG.

For the last experiments, we changed the incision closure technique, and, instead of a straight line, we performed the HM procedure. The HM [36] is a common, clinical procedure utilized for correcting pyloric stenosis or enlarging the diameter of narrowed bowel segments. It is performed by creating a longitudinal incision along the antimesenteric border of the small bowel and then closing it transversely [37]. It can assist obstructed patients to resume normal digestive functions and improve their quality of life. Hence, in this case, the robotic movement was also changed, and a vertical path movement was added.

*Burst pressure test*
After soldering or suturing, both sides of the sample were clamped. Then, we carefully inserted an injection needle 3-4 cm from the incision area to ensure the incision site remained unaffected. For the undamaged tissues, the needle was inserted at an arbitrary position. Before introducing water, we withdrew the needle, leaving only a flexible plastic tube inside. The tube was attached to a Harvard Apparatus syringe infusion pump model 22. Water was then introduced into the bowel lumen at a constant rate of 18 [ml/min]. For each sample, we recorded the burst pressure, defined as the pressure at which the tissue began leaking water either from the incision area or elsewhere in the bowel. For data analysis, we calculated the normalized burst pressure by dividing burst pressure for the sutured or soldered tissue by the median burst pressure value for the undamaged tissue group.

*Statistical analysis*
To compare the different robotic movement trajectories, we performed a two-way ANOVA model using the normalized burst pressure as the dependent variable. The incision closure technique (manual sutures or robotic soldering), the robotic movement trajectory (zigzag, fast discrete, slow discrete or the double slow discrete) and their interaction were used as the independent variables. When interaction was found to be significant, we used t-test for post hoc analysis with Bonferroni correction for multiple comparisons. Statistical significance was determined at the 0.05 threshold in all tests.

For the fourth and fifth experiments, in which different concentrations were examined, we performed a one-way ANOVA model with the normalized burst pressure used as the dependent variable. The incision-closing method (manual sutures or different robotic soldering concentrations) was used as the independent variable.

For comparing the two different days with the HM protocols, we performed a two-way ANOVA model in which the normalized burst pressure was used as the dependent variable. The incision closure technique (manual suturing or robotic soldering), the experiment location (BGU lab or Lahav facility), and their interaction were used as the independent variables.

**Table 1: Experimental protocols for the different parameters.**

| Experiment number | Number of soldered samples | Robotic movement parameters | | | | Laser tissue distance [mm] | Solder solution concentration | |
|---|---|---|---|---|---|---|---|---|
| | | trajectory | Distance between steps [mm] | Velocity [mm/sec] | Dwell time [sec] | | Albumin [mg/ml] | ICG [mg/ml] |
| 1 | 5 | Zigzag | Vertical:1.5 Horizontal: 1 | 0.5 | 12 | 7 | 800 | 0.6 |
| 2 | 4 | Discrete | 1.5 | 0.3 | 10 | 35 | 800 | 0.6 |
| 3 | 3 | Discrete | 2 | 0.1 | 10 | 35 | 800 | 0.6 |
| 4 | 5 | Double | 2 | 0.1 | 10 | 35 | 800 | 0.6 |
| | 5 | | | | | | 735 | |
| | 5 | | | | | | 670 | |
| 5 | 3 | Double | 2 | 0.1 | 10 | 35 | 800 | 0.6 |
| | 3 | | | | | | | 0.45 |
| | 3 | | | | | | | 0.3 |
| 6 | 4 | HM | 2 | 0.1 | 10 | 35 | 800 | 0.45 |
| 7 | 5 | HM* | 2 | 0.1 | 10 | 35 | 800 | 0.45 |

* The same experimental protocol as 6 at Lahav CRO facility.

*Cadaver experiment*

As our system is based on tissue temperature control during the procedure, we analyzed the changes required to be made in the temperature protocol prior to performing in-vivo experiments. The ex-vivo experiments were performed on tissues at room temperature ($25^oC$). However, the porcine body temperature is around $38^oC$, while the operation room temperature was approximately $14^oC$. Due to the low temperature in the operating room, the temperature set point was set to $50^oC$.

For this experiment, we used a fresh cadaver (weighing ~ 50-60 kg) euthanized half an hour prior to our procedure. For each section of the cadaver's small intestine, we performed the HM procedure as described in experimental setup number 6 and 7 (Table 1), according to the chosen protocol. We performed one incision for each sample area. Then, we applied three approximation sutures with Vicryl 4.0 and performed our procedure. After soldering, we clamped both sides of the bowel proximally and distally to the incision area. We tested for gross leakage from the incision area by filling this section with 20 ml of water. When we saw that water was not coming out of the incision area, we proceeded to perform the burst pressure test.

*In-vivo experiment protocol*

We performed the Heineke-Mikulicz procedure using the RLTS system on 6 pigs (weighing ~50-60 kg) in the Institute for Animal Research of Kibbutz Lahav. The experimental protocol was approved by the National Permit Committee for Animal Science, Israel, approval number

2212-385-4, dated December 18th, 2022. The experiment was performed on two separate days, two weeks apart. Throughout the entire procedure, the animals were anesthetized and ventilated. For each animal, we performed a small abdominal incision, exposed a small area of the small intestine, and performed one longitudinal incision, which was then closed with the HM procedure; 3 approximation sutures with Vicryl 4.0 were applied. Next, we performed the HM procedure using the solder solution of 800 mg/ml bovine albumin mixed with 0.45 mg/ml ICG. The temperature set point was established to $50^{\circ}C$ due to the low temperature in the operating room. We also had to compensate for the movements in live animals undergoing surgery: 1) The peristaltic movement of the intestine and 2) The breathing motion of the animal's chest. Due to both, the tip of our probe was raised to 45 mm above the sample and the feedback loop between the radiometer and the laser increased or reduced the intensity of the beam on the heated spot so that its temperature was constant. After completing our protocol, we tested for gross leakage from the incision area by filling a section of 10 cm in the incision area with 20 ml of water. We then added two non-absorbable marking sutures at a distance of ~10 cm proximal and distal to the incision so that we could find the soldering area after recovery. We followed the animal's activity for two weeks. We recorded their body temperature and their bowel activity by bowel movement and the time transitioning between liquid and solid food. After two weeks, the animals were euthanized and an approximately 10 cm long segment of the procedure area was removed and sent to histology. A qualitative histology test was conducted for samples from three areas of the tissue. The control group was taken from a tissue outside the soldering area and the mesenteric, and the antimesenteric were taken from a tissue in the soldering area. The histology test examined parameters for tissue fusion and healing process such as tissue regeneration as well as parameters for tissue damage such as thermal injuries.

**Funding**
This work was supported by
Ministry of Science and Technology, Israel, grant number 15627-3
Ben-Gurion University of the Negev through the Agricultural, Biological and Cognitive Robotics Initiative (funded by the Marcus Endowment Fund and the Helmsley Charitable Trust)
Kreitman Fellowship at Ben-Gurion University of the Negev, Israel.

**Author contributions**
Conceptualization: IN, AK, UN
Methodology: SA, BR, MP, SB, YS, IA, TP, IN, AK, UN
Investigation: SA, BR, MP, SB, YS, IA, TP, IN, AK, UN
Visualization: SA, BR
Funding acquisition: IN, AK, UN
Project administration: SA, IN, AK, UN
Supervision: IN, AK, UN
Writing – original draft: SA, MP, UN
Writing – review & editing: SA, BR, MP, SB, YS, IA, TP, IN, AK, UN

**Competing interests**
Authors declare that they have no competing interests.

**Data and materials availability**
https://github.com/Bio-Medical-Robotics-BGU/Robot-Laser-Tissue-Soldering